\documentstyle[aps,prl,preprint]{revtex}
\begin{document}
\draft
\title{ Raman Response in Doped Antiferromagnets }
\author{  P. Prelov\v sek and J. Jakli\v c }
\address{ J. Stefan Institute, University of Ljubljana, 
61111 Ljubljana, Slovenia }
\date{\today}
\maketitle
\begin{abstract}

The resonant part of the $B_{1g}$ electronic Raman scattering response
is calculated within the $t-J$ model on a planar lattice as a function
of temperature and hole doping, using a finite-temperature
diagonalization method for small systems. Results, directly applicable
to experiments on cuprates, reveal on doping a very pronounced
increase of the width of the two-magnon Raman peak, accompanied by a
decrease of the total intensity. At the same time the peak position
does not shift substantially in the underdoped regime.

\end{abstract}


%\narrowtext
Proper understanding of electronic properties of cuprates,
representing a remarkable example of strongly correlated systems,
remains a challenge for theoreticians and experimentalists. One of
very useful probes for the investigation of electronic excitations in
these materials has been the Raman scattering. Using the latter
method, it has been clearly established that in the reference
insulating substance, e.g. in $\mathrm{La_2CuO_4}$, the most
pronounced electronic Raman processes at low temperatures correspond
to the short-wavelength magnetic (two-magnon) excitations \cite{lyon},
which can be well described within the Heisenberg model for the planar
antiferromagnet (AFM) \cite{sing}.

A more general framework for the theoretical explanation of the Raman
scattering in strongly correlated systems, and specifically in
cuprates, has been first given within the Hubbard model, where the
effective Raman operator for resonant and off-resonant conditions has
been derived \cite{shas}.  Nevertheless, due to the difficulties in
performing the theoretical analysis and due to the possibility of
other relevant processes, an agreement on the appropriate
interpretation is still lacking for a number of Raman scattering
results in cuprates.

One aspect concerns the pronounced temperature dependence of the
linewidth in the undoped AFM \cite{knol}. The latter has been
attributed to the phonon-induced broadening \cite{nori}, but an
important role could be as well played by higher-order resonant
processes \cite{chub}, becoming relevant in the resonant-Raman
conditions.

Another problem is the doping dependence on the Raman scattering
\cite{lyon1,blum}. Recent experiments, performed on $\mathrm{YBaCuO}$
materials in the resonant regime \cite{blum}, show a dramatic
increase of the broadening of the two-magnon peak with doping, so that
spectra appear essentially flat in the normal phase $T>T_c$ when
approaching the `optimum' doping cases with highest $T_c$. At the same
time, the peak position does not move appreciably.

The aim of the present paper is to analyse the influence of doping and
finite temperatures on the Raman spectra within the framework
of the one-band $t-J$ model, assuming that the dominant resonant
contribution remains of the Loudon-Fleury type due to the spin
exchange \cite{fleu}. Assuming the Raman operator we calculate the
spectra at various hole concentrations $c_h$ and temperatures $T$ by
employing the recently introduced finite-T diagonalization method for
small correlated systems \cite{jakl}. $T>0$ results are of interest by
themselves. However, in the case of Raman spectra, where quite broad
features are expected particularly in doped systems, using finite
$T>0$ mainly represents a technical advantage to obtain
macroscopic-like spectra even for small systems, as already
established in previous applications of the method to the evaluation
of optical conductivity and dynamical spin correlations
\cite{jakl1}. The improvement over the usual $T=0$ Lanczos method 
can be judged by comparing our results with previous Raman spectra
obtained within the $t-J$ model containing few mobile holes
\cite{poil}. While some aspects appear similar (e.g. lowest frequency
moments), actual spectra calculated at $T=0$ from the ground state
are dominated by few peaks, which are clearly size-dependent.

One of the most studied prototype models of correlated systems, and
specifically of the low-energy properties of cuprates, is the $t-J$
model \cite{rice}
\begin{equation}
H=-t\sum_{\langle ij\rangle  s}(c^\dagger_{js}c_{is}+ \text{H.c.})
+J\sum_{\langle ij\rangle} (\vec S_i\cdot \vec S_j - 
{1\over 4} n_i n_j), \label{eq1}
\end{equation}
where $c^\dagger_{is}(c_{is})$ are projected fermionic operators,
prohibiting double occupancy of sites. The ground state of the model
has been studied both by analytical \cite{rice} and numerical methods
\cite{dago}. Recent investigations of the finite-$T$ dynamical spin
and charge response \cite{jakl1} established also qualitative (as well
as quantitative) contact with anomalous normal-state properties of
cuprates, confirming the $t-J$ model as the proper framework for
studies of the low-energy physics for such doped AFM.

The operator, relevant for the resonant Raman scattering in cuprates,
cannot be determined uniquely within the $t-J$ model (in contrast to
operators for some other charge and spin response functions), since
necessarily Raman processes involve higher resonant levels. Here we
adopt a view that more complete models for cuprates, e.g. the
three-band model, can be for processes of interest mapped onto an
effective $t-U$ Hubbard model \cite{hybe,rice}, where the upper
Hubbard-band states now correspond to charge transfer excitations in
cuprates. Within the Hubbard model near half-filling the
Raman-scattering operator has been derived in the limit $~t/U \ll 1$
\cite{shas}, yielding the well known form for the  Heisenberg AFM
\cite{fleu}
\begin{equation}
R_1= A \sum_{\langle ij\rangle} (\vec \epsilon_{inc}\cdot \vec r_{ij})
(\vec \epsilon_{sc}\cdot \vec r_{ij}) (\vec S_i\cdot \vec S_j - 
{1\over 4} n_i n_j), \label{eq2}
\end{equation}
where $\vec \epsilon_{inc}, \vec \epsilon_{sc}$ are the incident and
the scattered electric-field vector directions, respectively, $\vec
r_{ij}$ is the vector connecting sites $i$ and $j$, and the amplitude
factor $A = 4 t^2/(U-\omega_{inc}) \propto J$ incorporates the
resonance at the incident-light frequency $\omega_{inc} \sim U$.

For the undoped (insulating) case the operator $R_1$ is the only
resonant term of the order $t^2/U$, hence the dominant one, at least
outside the resonance $|U-\omega_{inc}|>t$. In the doped AFM
additional processes of the same order of magnitude are possible. The
latter correspond to the hopping term involving three neighboring
sites, leading partly also to the next-neighbor hopping correction,
i.e. to the $t^\prime$ term, within the $t-t^\prime-J$ model
\cite{rice,hybe},
\begin{equation}
R_2= {A\over 4} \sum_{(i,j,k), ss'} (\vec \epsilon_{inc}\cdot \vec
r_{ij}) (\vec \epsilon_{sc}\cdot \vec r_{jk}) c^\dagger_{ks'}c_{js'}
c^\dagger_{js}c_{is}. \label{eq3}      
\end{equation}
The contribution of such term clearly scales with the hole
concentration (it vanishes at half filling), so it is a quantitative
question whether it can become important in the relevant parameter
regime of cuprates.

The Raman spectral function is given by
\begin{eqnarray}
I(\omega) &=& {1\over \pi N}\text{Re}\int_0^\infty dt ~e^{i \omega t}
\langle R(t)R(0)\rangle = \nonumber\\
&=& {1\over N Z } \sum_{n, m} e^{-\beta E_n}
|\langle m|R|n\rangle|^2 \delta (\omega-E_m+E_n)~, \label{eq4}
\end{eqnarray}
where $\omega=\omega_{sc}-\omega_{inc}$, $\beta=1/T$ (we use furtheron
$k_B=\hbar=1$), and $Z$ is the partition function. Sums run over all
eigenstates $n,m$ with corresponding energies $E_n, E_m$ and are
size-normalized with $1/N$.

To calculate $I(\omega)$ within the $t-J$ model on a square lattice we
study finite clusters with $N=16 - 20$ sites. Via Eq.(\ref{eq4}) this
would be impossible with the method of full diagonalization
\cite{bacc}, used previously to investigate the Raman response for the
Heisenberg model with maximum lattice size $N=16$. Instead, we employ
the recently developed method for finite-$T$ dynamical (and statical)
correlation functions \cite{jakl}, based on the Lanczos iteration
procedure combined with random sampling. The method is quite
convenient for the analysis of the Raman response, in particular for
doped systems, since quite broad spectral features and a weak
$T$-dependence require a modest number of Lanczos steps $M < 100$
\cite{jakl,jakl1}. In results presented below we thus reach systems
with $N_{st} \sim 400.000$ states in the largest $\vec q, S_z$-basis
sector, while the sampling involves typically $N_0 \sim 200-500$
initial configurations. As discussed quite extensively in connection
with previous applications, the spectra reveal a macroscopic-like
behavior only at finite $T>T^*$, where $T^*$ is related to the
low-energy level spacing and thus dependent on the system size and
hole doping. In all calculations we will fix $J/t =0.3$, as relevant
for cuprates \cite{rice}, where $J \sim 950~\text{cm}^{-1}$. For
such parameters we typically reach $T^* \sim 0.1~t = J/3$ in the doped
system with $0.1<c_h<0.3$.

We restrict our analysis to the dominant $B_{1g}$ scattering geometry
with $\vec \epsilon_{inc} =(\vec e_x +\vec e_y)/\sqrt{2}$ and $\vec
\epsilon_{sc} = (\vec e_x -\vec e_y)/\sqrt{2}$.  Let us first discuss
results for the undoped Heisenberg model. Here the low-energy levels
are quite sparse and hence we reach only $T^* \sim 0.5~J$ in the
largest $N=20$ system. In Fig.~1 we present the $I(\omega)$ spectra
for several $T>T^*$. Consistent with $T \sim 0$ results
\cite{sing,nori} and with the low-$T$ experiments \cite{knol} we
observe a two-magnon peak at $\omega \sim 3.3~J$, being quite narrow
at low $T<J$, and having the width limited by quantum fluctuations
\cite{sing}. From Fig.~1 it follows that the peak width starts to
increase substantially only at higher $T \sim J$, where a gradual
transition to a broad featureless spectrum occurs. It should be also
noted that in spite of large broadening the peak does not move with
$T$ up to $T\sim 2~J$. Nevertheless, when considering undoped
cuprates, such as $\mathrm{La_2CuO_4}$ with $J \sim 1400~\text{K}$,
clearly other mechanisms have to be invoked to account for the
observed pronounced $T$-dependent width at lower $T\ll J$
\cite{nori,chub}.

We proceed to the doped systems, where we first consider only the
exchange part $R=R_1$, Eq.(\ref{eq2}). Here $T$ seems to play a less
essential role, provided that $T^*< T < J$. Typically we observe for
doped systems only a steady decrease of the intensity with $T$, e.g. a
$\sim 20\%$ reduction in the interval $T/J = 0.3 - 1.0$ for $c_h \sim
0.2$, and $\sim 30 \%$ for $c_h \sim 0.05$. On the other hand, the
dependence on doping is essential, as evident from Fig.~2, where we
present spectra for various hole concentrations $c_h=N_h/N \le 0.25$
at lowest $T=0.15~t > T^*$. Already the smallest possible (for
available sizes $N$) finite doping $c_h =0.05$ increases dramatically
the width of the two-magnon peak, and spectral features become
overdamped on approaching the `optimum' doping $c_h \sim 0.2$. It is
however remarkable, that the peak position does not shift appreciably
in the underdoped regime $c_h \alt 0.1$. Only for $c_h > 0.15$ the
spectra change to a broad central-peak form with a maximum at $\omega
= 0$.

Another experimentally relevant quantity is the total spectral
intensity $I_0 = \int_0^\infty I(\omega) d\omega $ and its variation
with doping.  At the same time, we evaluate the frequency moments
$\langle
\omega \rangle$, $\langle \omega^2 \rangle$,  and the spectral width
$\sigma=\sqrt{\langle\omega^2\rangle-\langle \omega\rangle^2}$, where
the averages are calculated with respect to the $I(\omega>0)$ part of
the spectra. Quantities on Fig.~3 are calculated within various-size
systems for $T=0.5~J$, where values should be quite close to the $T=0$
results.  Results for finite $c_h>0$ are presented for $N=16-20$. For
undoped system $N_h=0$ we include values for $N=20$, as well as for
$N=26$ at $T=0$. In the latter case, we get e.g. the width $\sigma
\sim 0.8~J$, very close to analytical predictions \cite{sing}. The
scattering of values in Fig.~3 for similar $c_h$ could be partly
attributed to the error due to the  restricted random sampling
\cite{jakl} for larger systems, while smaller systems at chosen 
$T\sim T^*$ could still suffer from finite size effects.

As evident from Fig.~3, the total intensity $I_0$ decreases steadily
but substantially with doping, i.e. by a factor $\sim 3$ from $c_h=0$
to $c_h =0.25$. The average frequency $\langle \omega \rangle$ starts
to increase slowly only at $c_h>0.15$. The enhancement can be
attributed to the emergence of high-frequency tails with $\omega>t$ in
doped systems, whereas the peak position moves in the opposite way. On
the other hand, the spectral width $\sigma$ shows a more dramatic
variation, in particular at low doping $c_h <0.1$.

Let us finally discuss the influence of the three-site hopping term
$R_2$, Eq.(\ref{eq3}). Our results indicate that it is less important
at low $T$ and in the concentration range of interest $c_h
\le 0.25$, either calculated separately as $R=R_2$ or combined 
$R=R_1+R_2$. When evaluated as $R=R_2$ we observe that $\Delta
I(\omega)$ is essentially featureless up to $\omega \sim 4~t$, and
becoming larger only at $T>J$, while at $T\sim T^*$ it amounts only to
e.g. $0.2~I_0$ at $c_h =2/16$.  Taken into account as $R=R_1+R_2$, the
largest corrections are at $c_h=0.25$ as expected, i.e. $\delta
I_0/I_0 \sim 0.3$. Nevertheless, the most pronounced changes are not
in the regime $\omega<2~t$ (spectra are even flatter there), but
rather in the large-$\omega$ tails, leading to substantially increased
$\langle
\omega \rangle$ (e.g. $\sim 70\%$ for $c_h \sim 0.25$) and $\sigma$ at
the `optimum' doping. It seems that the second-neighbor-hopping
processes are quite restricted at low $T$, due to remaining
short-range magnetic correlations, disappearing only at higher
$T>J$. It should be however noted that we did not introduce an
analogous three-site term (or a simpler $t^\prime$ hopping term) in
the $t-J$ model, Eq.(\ref{eq1}) \cite{rice,hybe}, which possibly could
lead to some quantitative modifications.

How can we interpret the above results for the $B_{1g}$ Raman
scattering? Up to the overdoped situation $c_h \sim 0.3$ the
scattering seems to be dominated by the spin-exchange part,
Eq.(\ref{eq2}), since the latter appears to determine the low-energy
fluctuations even at the `optimum' doping \cite{jakl1}. Still, there
are evident changes with doping. In an undoped AFM a well developed
longer-range order (long-range at $T=0$) induces a sharp peak in
$I(\omega)$. In the underdoped regime $c_h < 0.1$ the reduced AFM
correlation length does not shift the peak, but induces a large
broadening and a reduction of the intensity. In this respect the doped
system behaves quite similarly to an undoped AFM but at elevated
$T\sim J$, as one can conclude from the similarity of spectra on
Figs.~1,2. For $c_h >0.1$ the spin system seems to approach a totally
incoherent one. Any coherence in the contribution of different bonds
in Eq.(\ref{eq2}) is eliminated, leading to a reduction of the
intensity and a broad featureless spectrum with a central maximum at
$\omega=0$.

Finally let us comment on the relation of our results to
experiments. A systematic resonant-Raman scattering study has been
recently performed on a sequence of $\mathrm{YBa_2Cu_3O_{6+x}}$
materials and $\mathrm{YBa_2Cu_4O_8}$ \cite{blum}. Although it is not
straightforward to relate in this case the actual doping of
$\mathrm{CuO_2}$ layers with our model doping $c_h$ (and consider also
the possible role of $\mathrm{CuO}$ chains), it seems that experiments
for the underdoped materials correspond well to our model results,
both regarding the shape of the Raman spectra and their intensity
variation with doping. On the other hand, entering the
`optimum'-doping and the overdoped regime both experiments and the
theory show consistently a nearly flat Raman response.
              
One of the authors (P.P.) wants to acknowledge stimulating
discussions with M. Klein and G. Blumberg.

\begin{figure}
\caption{Raman intensity $I/A^2$ vs. $\omega/J$ for the undoped
Heisenberg model at various $T$, as calculated for $N=20$. An
additional smoothening width $\Delta = 0.2~J$ is used at low $T$.}
\end{figure}

\begin{figure}
\caption{$I/A^2$ vs. $\omega/J$ for the $t-J$ model at various
hole concentrations $c_h$ and $T=0.5~J$. The smoothening is $\Delta =
0.2~J$.}
\end{figure}

\begin{figure}
\caption{Total Raman intensity $I_0$ (triangles), the average 
frequency $\langle \omega\rangle /J$ (circles), and the spectral width
$\sigma/J$ (squares) vs. doping $c_h$ at $T=0.15~t$, as evaluated for
systems with $N=16-20$. For $c_h=0$ also $N=26$ results at $T=0$ are
included. Lines are guides to the eye only.}
\end{figure}

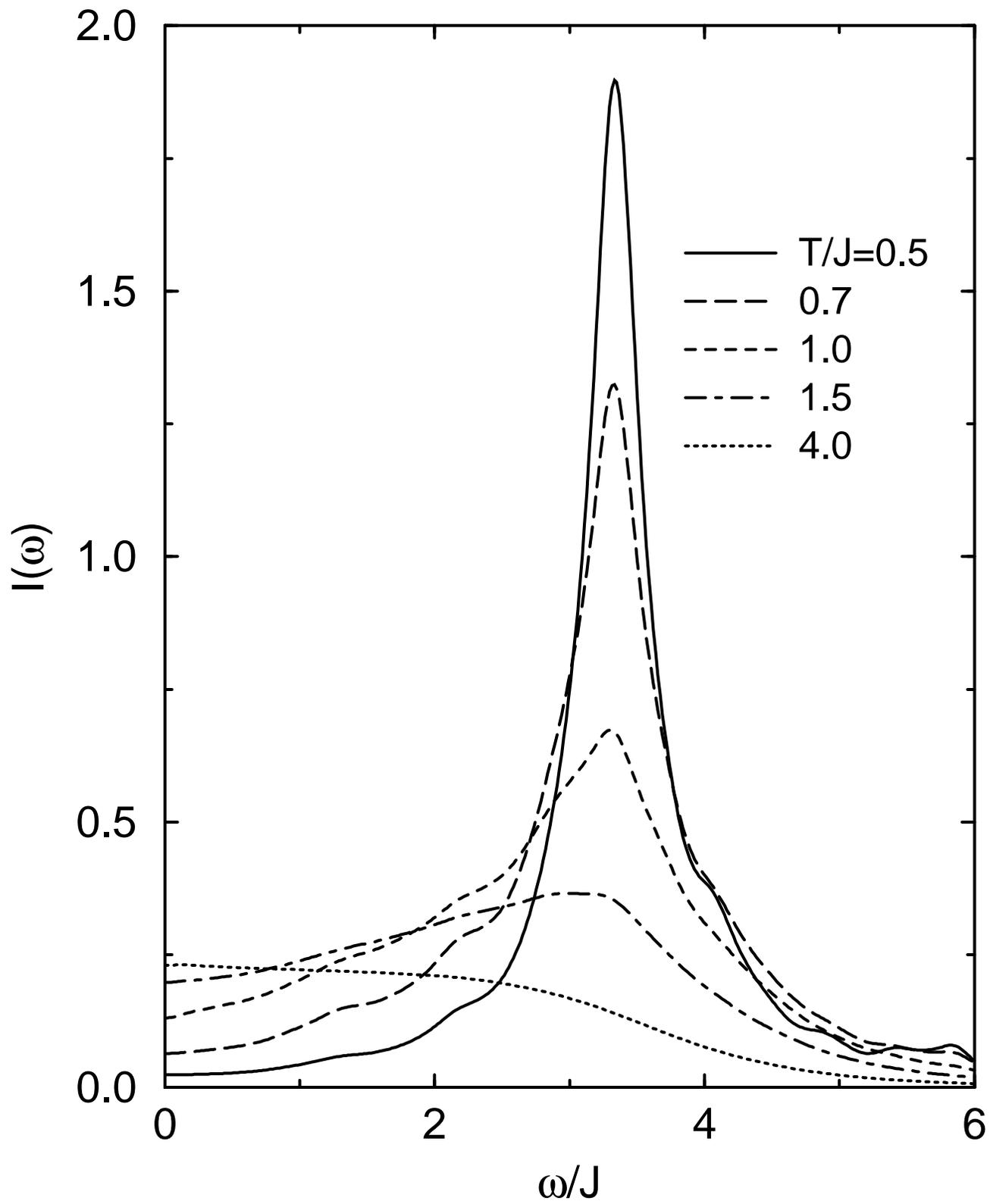
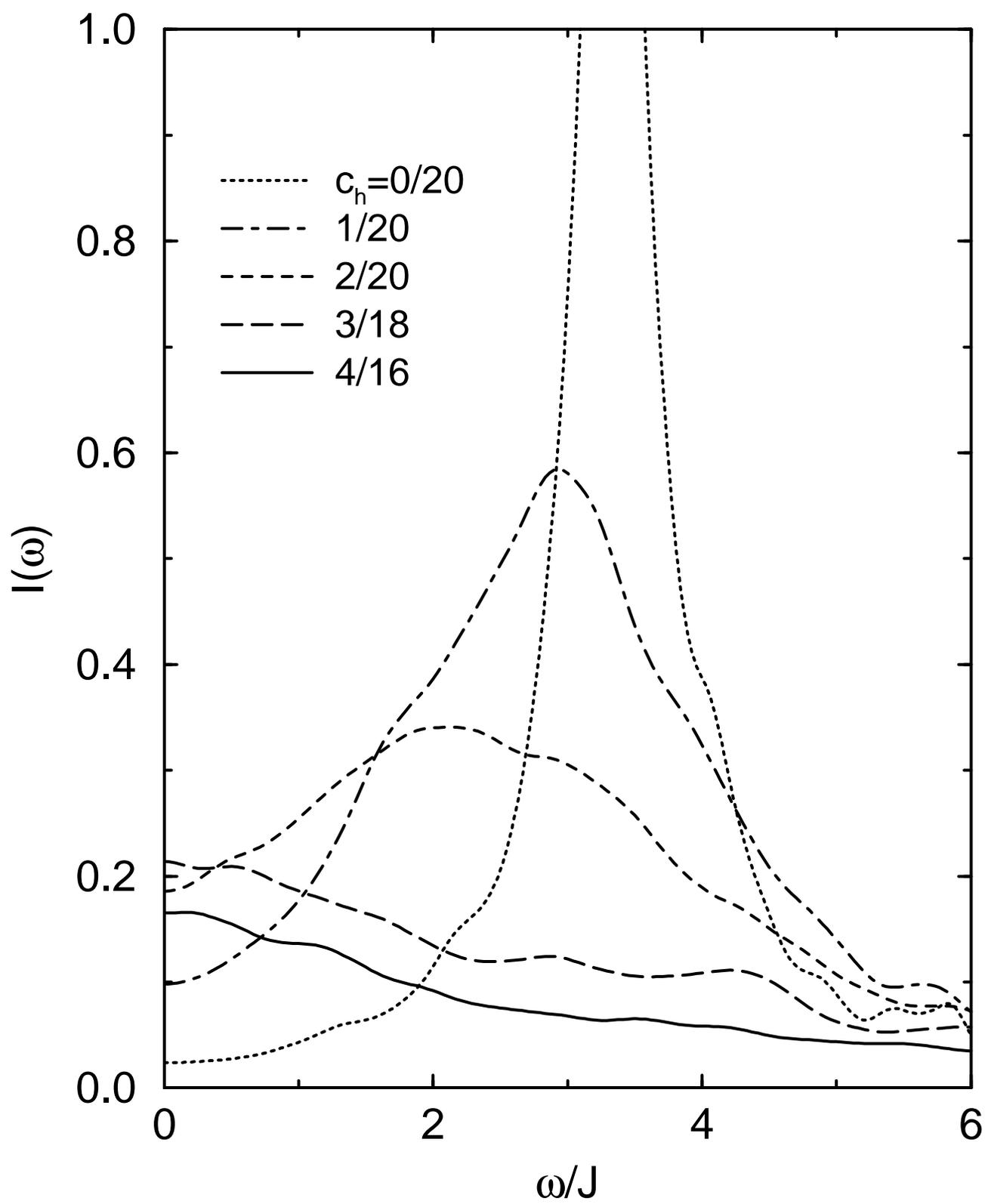
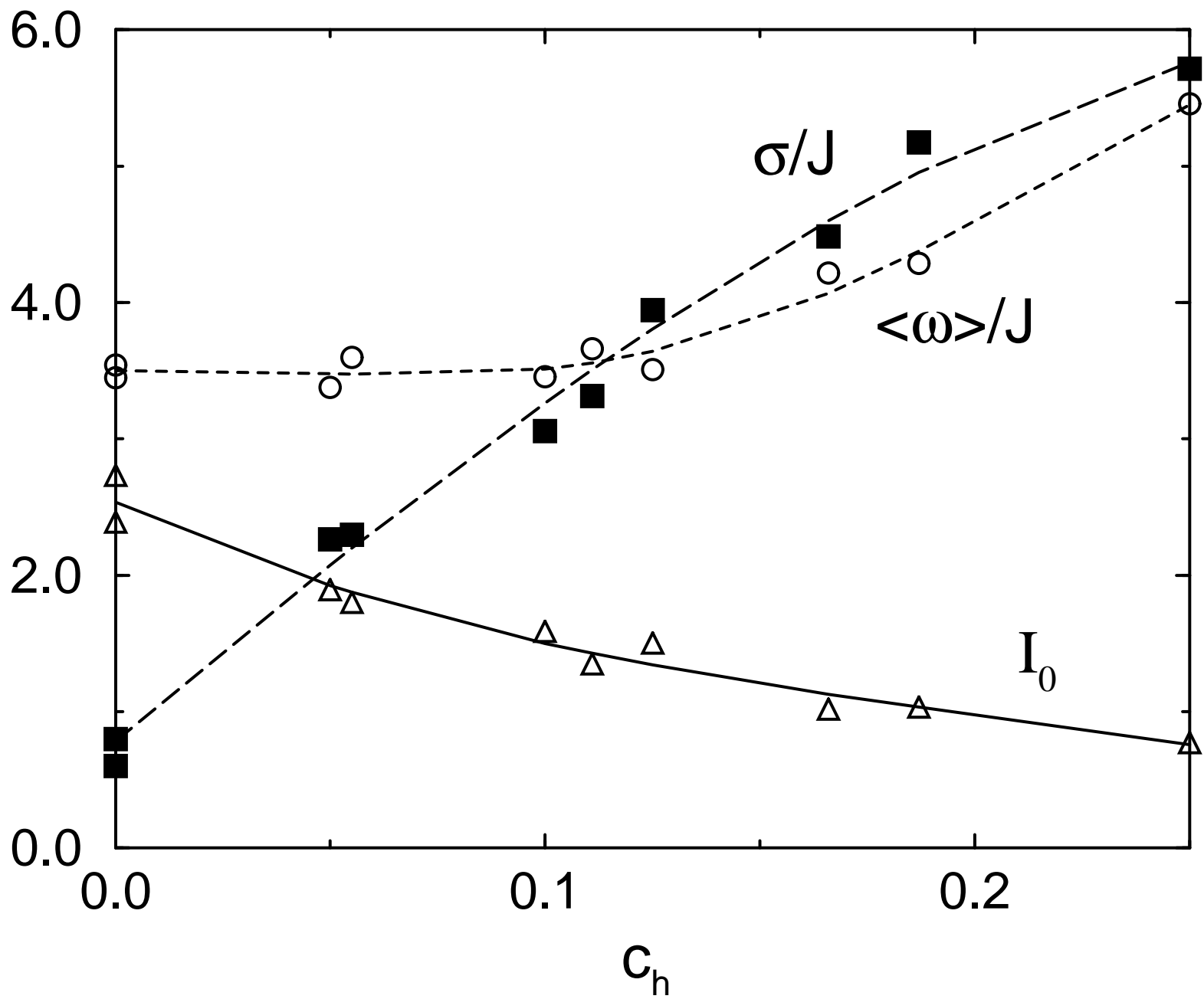
\end{document}